\documentclass[journal=jacsat,manuscript=letter]{achemso}
\usepackage{epsfig}
\usepackage{float}
\usepackage{color}
\usepackage{array}
\usepackage{amsmath}
\usepackage{amssymb}
\usepackage{lipsum}
\usepackage{epstopdf}
\usepackage[version=3]{mhchem}
\setkeys{acs}{usetitle = true}
\author{Jorge R. Espinosa $^{\dagger}$}
\author{Carlos Vega $^{\dagger}$}
\author{Eduardo Sanz $^{\dagger}$}
\email{esa01@quim.ucm.es}
\affiliation[Universidad Complutense de Madrid]
{$^{\dagger}$Departamento de Quimica Fisica I,
Facultad de Ciencias Quimicas, Universidad Complutense de Madrid,
28040 Madrid, Spain.\\}
\title[An \textsf{achemso} demo]
  {Homogeneous Ice Nucleation Rate in Water Droplets}

\begin{document}

\begin{abstract}

To predict the radiative forcing of clouds it is necessary to know the rate with which ice homogeneously nucleates in supercooled water. 
Such rate is often measured in drops to avoid the presence of impurities. 
At large supercooling small (nanoscopic) drops must be used to prevent simultaneous nucleation events.
The pressure inside such drops is larger than the atmospheric one by virtue of the Laplace equation. 
In this work, we take into account such pressure raise in order to predict the nucleation rate in droplets
using the TIP4P/Ice water model. We start from a recent estimate of the maximum drop size that can be 
	used at each supercooling avoiding simultaneous nucleation events [Espinosa et al. J. Chem. Phys.,  2016]. 
	We then evaluate the pressure inside the drops 
	with the Laplace equation. Finally, we obtain the rate as a function of the supercooling by   
	interpolating our previous results for 1 and 2000 bar [Espinosa et al. Phys. Rev. Lett. 2016] using the Classical Nucleation Theory 
	expression for the rate. 
	This requires, in turn, interpolating the ice-water interfacial
	free energy and chemical potential difference. The TIP4P/Ice rate curve thus obtained is in good agreement 
	with most droplet-based experiments. In particular, we find a good agreement with measurements performed 
	using nanoscopic drops, that are currently under debate. The successful comparison between model and experiments
	suggests that TIP4P/Ice is a reliable model to study the water-to-ice transition and that Classical Nucleation 
	Theory is a good framework to understand it. 

\end{abstract}

To make climate change predictions it is necessary to estimate the radiative forcing (the balance between absorbed and reflected
solar radiation) caused by different factors. 
According to reports by the Intergovernmental Panel on Climate Change (IPCC), 
there are large uncertainties in the radiative forcing caused by clouds. Such 
uncertainty is partly due to the 
lack of reliable predictions of the ice content in clouds \cite{review_ice_formation_clouds_2005,baker97,demott10}. 
These predictions rely on estimates of the ice nucleation rate, $J$, or the number of
ice embryos that proliferate per unit time and volume \cite{baker97,review_ice_formation_clouds_2005,demott10}. 

In this paper we focus on the rate of homogeneous ice nucleation from pure water, $J_{hom}$. 
Although ice formation in the atmosphere is thought to occur predominantly heterogeneously from aqueous solutions \cite{review_ice_formation_clouds_2005,reviewMurrayheterogeneous2012},
the fact that clouds have been observed to supercool to very low temperatures (even below -35 $^o$C) \cite{de2011evidence,choi2010space,westbrook2011evidence,rosenfeld2000deep} 
suggests that there is homogeneous ice nucleation 
from nearly pure water in clean atmospheric conditions (upper troposphere). Moreover, 
ice nucleation from solution and heterogeneous ice nucleation are often treated 
as a sophistication of the case of homogeneous ice nucleation from pure water \cite{rasmussen1982ice,koopNature2000,turnbull1950kinetics}. 
It is therefore relevant to 
fully understand and characterise the latter. Of course, predicting the freezing of clouds requires knowledge
not only of the nucleation stage but also of the growth one. However, both freezing stages are sufficiently complex so 
as to deserve separate attention.

\begin{figure}[!htb]
\begin{center}
	\includegraphics[clip,width=0.8\textwidth]{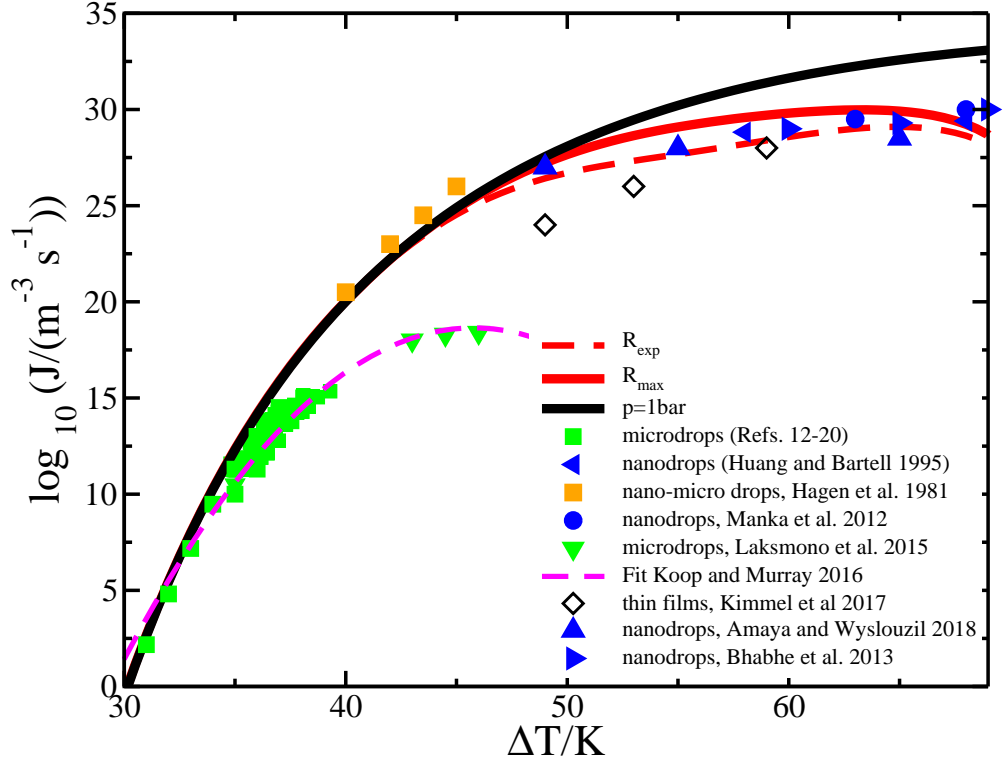}
	\caption{Ice nucleation rate at a function of supercooling, $\Delta T$ (difference between melting temperature
	and temperature of interest). Solid symbols correspond to experimental measurements in drops. 
	Green squares, correspond to micron sized drops from Refs. \cite{pruppacher1995,murray2010,riechers13,stockelJPCA2005,stanLabChip2009,kramer:6521,duft2004laboratory,leisnerPCCP2012,benz2005t}; downward green triangles are also micron sized
	drops but from Ref. \cite{laksmonoJPCL2015}; blue symbols correspond to nanoscopic drops \cite{manka2012,huang_bartell,bhabheJPCA2013,amaya2018ice}; orange symbols correspond to 
	drops in between the nanoscopic and the microscopic regime \cite{hagenJAS1981}. 
	Empty symbols correspond to measurements of the nucleation rate in thin films \cite{xu2017homogeneous}. 
	Dashed pink curve is a fit proposed in Ref. \cite{koopmurray2016} inspired
	by the publication of the data represented by the downward green triangles. Solid lines correspond to simulation estimates
	using the TIP4P/Ice water model obtained with Seeding \cite{seedingvienes}. The black line corresponds to the rate estimate at 1 bar 
	(from Ref. \cite{espinosaJCP2016}). In the red curves (this work) the effect of the 
	Laplace pressure inside the drops is taken into account in the simulation rate estimate. \textcolor{black}{Solid red corresponds
	to the rate measured in the largest possible drop where there is a single nucleation event, while dashed red corresponds to the rate in drops
	of size typically used in experiments}.}
\label{rates}
\end{center}
\end{figure}

Experiments to measure $J_{hom}$ typically use suspended droplets ranging from microscopic to nanoscopic size 
to avoid heterogeneous ice nucleation on impurities. 
In Fig. \ref{rates} $J_{hom}$ measurements as a function of the 
supercooling $\Delta T$ --the melting temperature minus temperature of interest-- are reported.
Green and blue symbols correspond to 
measurements performed with microscopic \cite{pruppacher1995,murray2010,riechers13,stockelJPCA2005,stanLabChip2009,kramer:6521,duft2004laboratory,leisnerPCCP2012,benz2005t} and nanoscopic \cite{manka2012,huang_bartell,bhabheJPCA2013,amaya2018ice} droplets respectively, while orange ones \cite{hagenJAS1981} 
correspond to droplet sizes in between both ranges.  
Recent measurements from 2015 \cite{laksmonoJPCL2015}, downward green triangles, inspired in 2016 a new fit to $J_{hom}$ 
(dashed pink curve \cite{koopmurray2016}) that
shows a maximum at $\Delta T$ $\sim$ 46 K. 
Such fit strongly clashes with measurements performed using nanoscopic drops at deep supercooling (blue points) \cite{manka2012,huang_bartell,bhabheJPCA2013,amaya2018ice}. 
According to these experiments $J_{hom}$ monotonously increases with supercooling, at least up to $\Delta T =$  70 K. 
Clarifying such discrepancy is a very relevant issue to atmospheric and climate science 
for the reasons explained in the 
previous paragraph. Several hypothesis have been put forward to explain the discrepancy \cite{laksmonoJPCL2015}. A plausible one is a spurious 
overestimation of the rate in nanoscopic drops due to nucleation at the air-water interface, 
but this remains a controversial issue \cite{haji2017perspective}.

\begin{figure}[!htb]
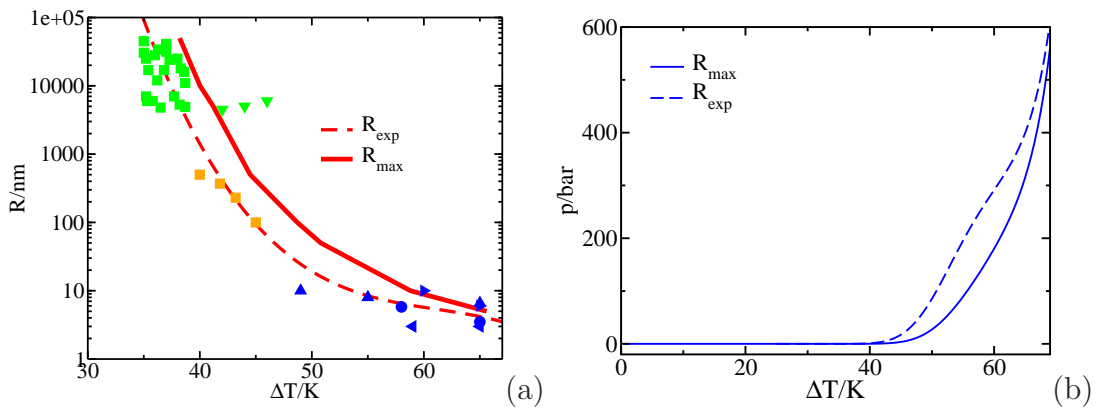

\begin{center}
	\includegraphics[clip,width=0.4\textwidth]{./figura_13_12june_apendice.eps}(a)
	\includegraphics[clip,width=0.4\textwidth]{./presion_vs_deltaT.eps}(b)
	\caption{(a) Droplet radius as a function of the supercooling. 
	The \textcolor{black}{solid red} curve corresponds to the maximum radius that enables measuring ice nucleation rates
	avoiding the simultaneous growth of several nuclei, $R_{max}$ \cite{espinosaJCP2016}.
	\textcolor{black}{The red dashed curve, $R_{exp}$, corresponds to a fit to the experimental data, excluding those
	given by downward green triangles}.
	Symbols correspond to the experiments indicated in the legend of Fig. \ref{rates}. 
	(b) Solid \textcolor{black}{(dashed)}: Laplace pressure inside drops of radius $R_{max}$ \textcolor{black}{($R_{exp}$)} 
	as a function of the supercooling.}
\label{sizes}
\end{center}
\end{figure}

In a recent work we have used TIP4P/Ice, a simple yet realistic water model \cite{JCP_2005_122_234511},
to predict $J_{hom}$ with computer simulations \cite{espinosaJCP2016,espinosaPRL2016} using the Seeding \cite{seedingvienes} and
the Mold Integration techniques \cite{espinosaJCP2014_2}. Our results for 1 bar, 
shown with a black curve in the figure, 
are in better agreement with the scenario supported by the nanoscopic drops measurements.
\textcolor{black}{In Ref. \cite{espinosaJCP2016} we argue that the measurements corresponding to the downward triangles could be 
underestimated because the employed drops may be too large 
and contain many ice nuclei simultaneously growing.}
Then, the time needed to observe water freezing would no longer be limited by 
the nucleation stage, but rather by the time required for the nucleated ice 
embryos to grow and fill a fraction of the drop volume that enables freezing detection. 
Since the rate is determined under the expectation that 
only one ice cluster nucleates in each drop \cite{laksmonoJPCL2015}, multiple nucleation events would 
lead to an underestimate of the nucleation rate. 
This  \textcolor{black}{multiple nucleation} aggravates as the supercooling increases because
$J_{hom}$ goes up with $\Delta T$. 
In Fig. \ref{sizes}(a) we reproduce our results from Ref. \cite{espinosaJCP2016} where,
by combining simulation estimates of $J_{hom}$ and of the speed of ice growth, 
we predicted $R_{max} (\Delta T)$, the maximum droplet radius that enables staying  
in the regime where 
drops are observed to freeze at the time required to 
nucleate a single critical ice cluster. As expected, $R_{max}$ goes down with $\Delta T$. 
Symbols in Fig. \ref{sizes}(a) have
the same legend as in Fig. \ref{rates}. Downward triangles, that inspired the
dashed pink fit in Fig. \ref{rates}, lie in the region where our simulations predict that many ice
clusters will simultaneously grow in the droplet.

We can now use $R_{max}(\Delta T)$ in conjunction with the Laplace equation, $\Delta P = 2\gamma_{lv}/R_{max}$, 
to estimate the pressure inside the largest drops that can be used if simultaneous nucleation events are to be avoided ($\gamma_{lv}$ is
the liquid-vapor surface tension).
To do such estimate we have used the $\gamma_{lv}$ temperature dependence given in Ref. \cite{hruby2014surface}, which 
we 
linearly extrapolated outside the reported measurement range (below -25$^o$C).  
The smooth variation of $\gamma_{lv}$ with temperature justifies such extrapolation. 
\textcolor{black}{Using $\gamma_{lv}$ for a flat interface could be inapropriate when dealing 
with curved drop surfaces. However, using a Tolman length of 1\AA --larger than the values typically reported \cite{joswiak2016energetic,lau2015surface,lau2015water}--
to correct for curvature effects only yields changes of less than 2 mN/m for the smallest drops considered. We therefore neglect 
any
curvature effects in $\gamma_{lv}$}.
The results for the pressure inside the drops as a function the corresponding supercooling are shown in Fig. \ref{sizes}(b), solid curve. 
For supercooling larger than $\sim$ 50 K the pressure sharply goes up. Therefore, rate measurements
using drops 
can no longer be performed at 1 bar for $\Delta T > 50 K$, which is an interesting conclusion of our analysis. 
This has to be taken into account when comparing simulation estimates with droplet based experimental measurements
of the nucleation rate. 
This issue has been disregarded in the black curve shown in Fig. \ref{rates}, which entirely corresponds to 1 bar  
(in simulations the rate is not computed inside drops but in the bulk thanks to 
periodic boundary conditions). The main aim of this paper is to 
provide a simulation prediction of $J_{hom}(\Delta T)$  
that can be directly compared with drop based measurements. 
This has been recently attempted in an experimental work, but only rough estimates were provided \cite{amaya2018ice}.

\begin{figure}[!htb]
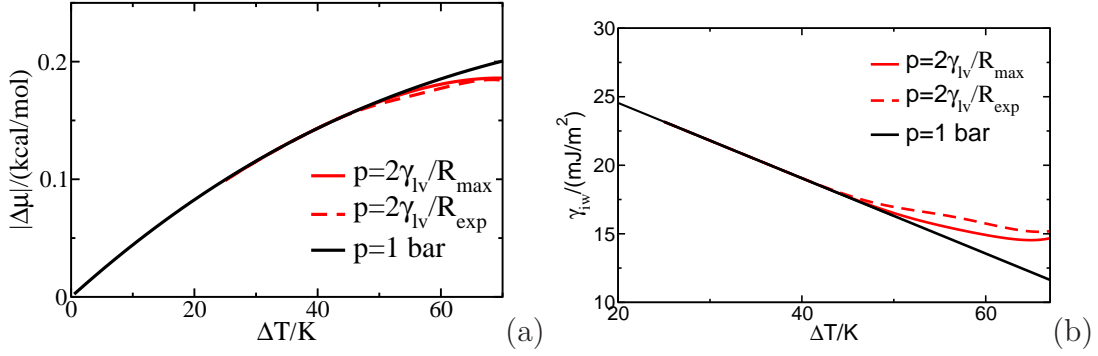

\begin{center}
	\includegraphics[clip,width=0.4\textwidth]{./deltamu.eps}(a)
	\includegraphics[clip,width=0.4\textwidth]{./gammastatisticalonlywithMIok.eps}(b)
	\caption{TIP4P/Ice predictions for the chemical potential difference between water and ice, (a), and the ice-water interfacial free energy, 
	(b), as a function of supercooling. Black curves correspond to 1 bar \cite{espinosaJCP2016}, solid red corresponds to the largest drops that can be used while avoiding simultaneous nucleation events \textcolor{black}{, and dashed red to drops with sizes typically used in the experiments. In order to parametrise
	the nucleation rate we use the following fits for the red solid curves: $|\Delta \mu|$ = 0.0012522 + 0.0044213 $\Delta T$ - 1.6401 $\cdot 10^{-5}$ $\Delta T^2$ - 1.259$\cdot 10^{-7}$ $\Delta T^3$ and $\gamma_{iw}$ = 30.157 - 0.3219 $\Delta T$ + 0.0042643 $\Delta T^2$ - 0.0001333 $\Delta T^3$ + 1.3504 $\cdot 10^{-6}$ $\Delta T^4$ and these for the red dashed ones: 
	$|\Delta \mu|$=0.00035032+0.0046013$\Delta T$-2.3187$\cdot 10^{-5}\Delta T^2$-6.9536$\cdot 10^{-8}\Delta T^3$ and
	$\gamma_{iw}$=29.986-0.25559$\Delta T$-0.0010465$\Delta T^2$+4.6503$\cdot 10^{-6}\Delta T^3$+2.9065$\cdot 10^{-7}\Delta T^4$.
	}}
\label{factors}
\end{center}
\end{figure}

To achieve this goal, one needs to compute for each $\Delta T$ the rate at the pressure given 
by $p(\Delta T)$ in Fig. \ref{sizes}. We have recently published $J_{hom}$ at 1 and 2000 bar for TIP4P/Ice \cite{espinosaPRL2016}. Here,
we interpolate our results to obtain $J_{hom}$ at the 
desired pressure. We compute $J_{hom}$ by plugging parameters obtained by simulations into
the expressions given by Classical Nucleation Theory (CNT) \cite{gibbsCNT1,gibbsCNT2,ZPC_1926_119_277_nolotengo,becker-doring}, a combination 
we call Seeding \cite{seedingvienes}. The CNT rate is given by:

\begin{equation}
	J_{hom}=A \exp\left(-\frac{C\gamma_{iw}^3}{K_BT\rho_s^2|\Delta \mu|^2}\right)
	\label{rate}
\end{equation}

Where $C$ is a constant that depends on the shape of the critical nucleus (here, 16$\pi$/3 for 
spherical clusters), $A$ is a kinetic pre-factor, $k_B$ is the Boltzmann constant, 
$\rho_s$ is the solid density, $\gamma_{iw}$ is the ice-water interfacial free energy and $|\Delta \mu|$
is the chemical potential difference between the bulk water and ice phases. 
\textcolor{black}{$|\Delta \mu|$ is computed with thermodynamic integration \cite{frenkelsmit,espinosaJCP2016} 
and $\gamma_{iw}$ with Mold Integration \cite{espinosaJCP2014_2} and Seeding \cite{seedingvienes} for
coexistence \cite{pocillosagua} and supercooled conditions \cite{jacs2013,espinosaJCP2016} respectively. 
The $\gamma_{iw}$ thus obtained has proven to give correct values for the nucleation rate 
when combined with Classical Nucleation Theory \cite{seedingvienes}. Therefore, the $\gamma_{iw}$ we use for
spherical critical clusters at supercooled conditions implicitly includes curvature and temperature corrections to that 
of a flat interface at coexistence}.

As shown in Fig. 3 of Ref. \cite{espinosaPRL2016}, neither $A$ nor the solid density significantly change with pressure. 
Therefore, we 
use the values of $A$ and $\rho_s$ corresponding to 1 bar for any supercooling.  
\textcolor{black}{To parametrise the rate we use the following fits for $A$ and $\rho_s$: 
$\ln$ ($A$/(m$^{-3}$ s$^{-1}$)) $= 91.656 - 0.11729 \Delta T - 0.00081401 \Delta T^2$
; $\rho_s$/(g/cm$^3$) $= 0.906+0.14 \cdot 10^{-3} \Delta T$}.

The chemical potential difference can be easily obtained
by thermodynamic integration from coexistence \cite{frenkelbook,0953-8984-20-15-153101}. 
In Ref. \cite{espinosaPRL2016} we showed that  $|\Delta \mu|$ does not
strongly change from 1 to 2000 bar. The smooth variation of $|\Delta \mu|$
with pressure enables us to obtain it by interpolation at the required pressure for each 
supercooling. 
The results are shown in Fig. \ref{factors}(a), where we compare $\Delta \mu (\Delta T)$ at 1 bar (black curve) with
that at the pressure given by $p(\Delta T)$ in Fig. \ref{factors}(b). Both curves are obviously the same
up to $\Delta T \sim 50 K$ where, according to Fig. \ref{sizes}(b), the pressure inside the drops is the atmospheric one. 
Beyond that supercooling $|\Delta \mu|$ is lower for the drops, which will contribute to lower $J_{hom}$ 
with respect to the bulk value ($|\Delta \mu|$ goes in the denominator of the exponential in Eq. \ref{rate}). 

We can also interpolate $\gamma_{iw}$ between our previously published values for 1 and 2000 bar \cite{espinosaPRL2016}.
The results are shown in Fig. \ref{factors}(b). Again, there is a noticeable effect at large supercooling:
$\gamma_{iw}$ increases due to the fact that, in virtue of the Laplace equation, the pressure inside
the drops exceeds the atmospheric one (as we have recently shown \cite{espinosaPRL2016}, the ice-water interfacial
free energy increases with pressure). From Eq. \ref{rate} it is clear that an increase of $\gamma_{iw}$ entails a 
decrease of the nucleation  rate.

Then, both $|\Delta \mu|$ and $\gamma_{iw}$ contribute to lower the nucleation rate inside the drops. 
With Eq. \ref{rate}, the red curves in Fig.\ref{factors} and the kinetic pre-factor previously obtained 
\cite{espinosaPRL2016} we can correct the black curve in Fig. \ref{rates} to account 
for Laplace pressure effects. The result is the red curve in Fig. \ref{rates}, which is now 
in very good agreement with nanoscopic drop data (blue symbols). In fact, the red curve is in good agreement 
with all drop-based rate measurements (solid symbols in Fig. \ref{rates}) except from those that 
inspired the fit with a maximum at $\Delta T = 46 K$ (dashed pink). 
\textcolor{black}{To obtain the solid red curve in Fig. \ref{rates} one needs
to combine in Eq. \ref{rate} the fits to the solid red curves of $\Delta \mu$ and $\gamma_{iw}$ given in the caption to Fig. \ref{factors} with 
those to $A$ and $\rho_s$ given above.}

\textcolor{black}{The red curve in Fig. \ref{rates} corresponds to the nucleation rate 
in the largest possible drop that can be used for each supercooling avoiding
multiple nucleation events (one with radius $R_{max}(\Delta T)$). 
However, the droplets employed in experiments need not 
be of radius $R_{max}$. In fact, in Fig. \ref{sizes}(a) one can see that 
the experimental droplet sizes typically lie below $R_{max}$. 
It is therefore interesting to compute the nucleation rate for the droplet sizes typically used 
in the experiments, given by a radius $R_{exp}$. We estimate $R_{exp}(\Delta T)$ by fitting 
the experimental values given in Fig. \ref{sizes}(a), excluding the downward green triangles
because they lie in the multiple nucleation events region. The $R_{exp}(\Delta T)$ fit is given by the dashed red curve 
in the figure. 
Given that $R_{exp} < R_{max}$, the 
pressure inside drops of radius $R_{exp}$ is larger than that inside drops of radius $R_{max}$ (see Fig. \ref{sizes} (b)).
In fact, as shown in Fig. \ref{sizes} (b), now the pressure departs from the atmospheric one at milder supercooling, 
$\Delta T = 45$ instead of $50$ K. 
Since a larger pressure causes a lower nucleation rate \cite{kannoScience1975,espinosaPRL2016}, 
the red dashed curve in Fig. \ref{rates}, corresponding to $R_{exp}$, lies below the 
solid red one, corresponding to drops with radius $R_{max}$. In fact, the $R_{max}$ curve in Fig. \ref{rates}
is an estimate of the highest possible rate that can be measured using drops and avoiding multiple nucleation events. 
The  $R_{exp}$ curve in Fig. \ref{rates} fits the experiments even better than the $R_{max}$ one, which 
further supports the reliability of the predictions given by our model.  
To obtain the dashed red curve in Fig. \ref{rates} one needs
to combine in Eq. \ref{rate} the fits to the dashed red curves of $\Delta \mu$ and $\gamma_{iw}$ 
given in the caption to Fig. \ref{factors} with 
those to $A$ and $\rho_s$ given above.
}

The experiments with thin films \cite{xu2017homogeneous} (empty diamonds in Fig. \ref{rates}) are carried out 
at atmospheric pressure (with a flat 
air-water interface). Therefore, they should be compared with the simulation predictions for 1 bar, black 
line in Fig. \ref{rates}. The comparison is not entirely satisfactory and further work is required to clarify
this issue. Furthermore, the comparison of thin film experiments with droplet experiments in a supercooling regime where 
drops are expected to be at nearly atmospheric pressure ($\Delta T <$ 45 K) does not look satisfactory either (as discussed in this work, 
for supercooling larger than 50 K thin film and droplet
experiments cannot be compared because the latter are carried out at a higher pressure). 

In summary, we have recently argued that there is a maximum droplet size that can be used at each supercooling
to measure the rate without having many ice nuclei simultaneously growing. Such size goes down with supercooling and, 
for supercooling larger than $\sim$ 50 K, the pressure inside the drop departs from the atmospheric one due to 
curvature effects (Laplace pressure). When the pressure of the liquid where ice nucleates increases, the nucleation 
rate decreases, mainly due to an increase of the interfacial free energy \cite{espinosaPRL2016}. Taking this into account we
provide simulation estimates of the homogeneous nucleation rate in droplets and we find a good agreement
with most droplet-based experimental measurements in a wide supercooling range. 
\textcolor{black}{The agreement is even better if drops with radius typically used in the experiments are considered (in this case a Laplace pressure
correction to the rate is noticeable for supercooling larger than 45 K)}.
Such a good agreement has several 
implications: (i) the data obtained at deep supercooling using nanoscopic drops (blue point in Fig. \ref{rates}) are supported by our simulations, while
those recently obtained with microscopic drops (downward green triangles in Fig. \ref{rates}) that inspired a fit to the nucleation rate with a maximum at a supercooling of 46 K (dashed pink line in Fig. \ref{rates}) are not; (ii)
TIP4P/Ice seems to be a good model to investigate both the thermodynamics and the kinetics of the water-to-ice transition; (iii) Classical Nucleation Theory
seems to be a solid framework to understand ice nucleation.

\textbf{Acknowledgements}

This work was funded by grants FIS2013/43209-P and FIS2016/78117-P of the MEC.
J.  R. Espinosa acknowledges financial support
from the FPI grant BES-2014-067625.  
The authors acknowledge the computer resources and technical assistance
provided by the Centro de Supercomputacion y Visualizacion de Madrid (CeSViMa).

\providecommand{\latin}[1]{#1}
\providecommand*\mcitethebibliography{\thebibliography}
\csname @ifundefined\endcsname{endmcitethebibliography}
  {\let\endmcitethebibliography\endthebibliography}{}


\begin{tocentry}
\includegraphics[clip=]{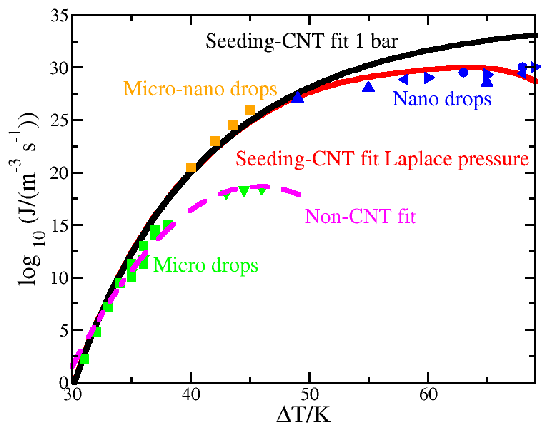}
\end{tocentry}

\end{document}